\documentclass[%
 aip,
 amsmath,amssymb,
 reprint,%
]{revtex4-1}

\usepackage{graphicx}
\usepackage{dcolumn}
\usepackage{bm}

\usepackage[utf8]{inputenc}
\usepackage[T1]{fontenc}
\usepackage{mathptmx}
\usepackage{float}

\begin{document}

\preprint{AIP/123-QED}

\title[All-optical naked-eye ghost imaging]{All-optical naked-eye ghost imaging\\}

\author{Gao Wang}
\author{Huaibin Zheng}%
 \email{huaibinzheng@xjtu.edu.cn}
\affiliation{ 
Electronic Materials Research Laboratory, Key Laboratory of the Ministry of Education \& International Center for Dielectric Research, School of Electronic and Information Engineering, Xi'an Jiaotong University, Xi'an 710049, China
}%
\author{Yu Zhou}%
\affiliation{ 
MOE Key Laboratory for Nonequilibrium Synthesis and Modulation of Condensed Matter, Department of Applied Physics, Xi'an Jiaotong University, Xi'an 710049, China
}%
\author{Hui Chen}%
\affiliation{ 
Electronic Materials Research Laboratory, Key Laboratory of the Ministry of Education \& International Center for Dielectric Research, School of Electronic and Information Engineering, Xi'an Jiaotong University, Xi'an 710049, China
}%
\author{Jianbin Liu}%
\affiliation{ 
Electronic Materials Research Laboratory, Key Laboratory of the Ministry of Education \& International Center for Dielectric Research, School of Electronic and Information Engineering, Xi'an Jiaotong University, Xi'an 710049, China
}%
\author{Yuchen He}%
\affiliation{ 
Electronic Materials Research Laboratory, Key Laboratory of the Ministry of Education \& International Center for Dielectric Research, School of Electronic and Information Engineering, Xi'an Jiaotong University, Xi'an 710049, China
}%
\author{Yuan Yuan}%
\affiliation{ 
Shaanxi Key Laboratory of Environment and Control for Flight Vehicle, Xi'an Jiaotong University, Xi'an 710049, China
}%
\author{Fuli Li}%
\affiliation{ 
MOE Key Laboratory for Nonequilibrium Synthesis and Modulation of Condensed Matter, Department of Applied Physics, Xi'an Jiaotong University, Xi'an 710049, China
}%
\author{Zhuo Xu}%
\affiliation{ 
Electronic Materials Research Laboratory, Key Laboratory of the Ministry of Education \& International Center for Dielectric Research, School of Electronic and Information Engineering, Xi'an Jiaotong University, Xi'an 710049, China
}%

\date{\today}

\begin{abstract}
Ghost imaging is usually based on optoelectronic process and eletronic computing. We here propose a new ghost imaging scheme, which avoids any optoelectronic or electronic process. Instead, the proposed scheme exploits all-optical correlation via the light-light interaction and the vision persistence effect to generate images observed by naked eyes. To realize high contrast naked-eye ghost imaging, a special pattern-scanning architecture on a low-speed light-modulation disk is designed, which also enables high-resolution imaging with lower-order Hadamard vectors and boosts the imaging speed. With this approach, we realize high-contrast real-time ghost imaging for moving colored objects.
\end{abstract}

\maketitle

\section{\label{sec:level1}Introduction}

More and more attention has been paid to ghost imaging due to its novel physical peculiarities and its potential applications in practice. The original ghost imaging experiments consisted of two correlated optical beams propagating in distinct paths and impinging on two spatially-separated photodetectors: the signal beam interacts with the object and then is received by a single-pixel (bucket) detector without spatial resolution, whereas the reference beam goes through an independent path and impinges on a spatial distribution detector, like charge-coupled device (CCD) without interacting with the object. Neither the bucket detector nor CCD can reveal the image of the object alone. One can retrieve the image by cross-correlating signals from bucket detector and CCD.
The first ghost imaging experiment was demonstrated by Pittman \textit{et\ al}\cite{RN9} in 1995 using entangled photon pairs. 
About ten years later, it was implemented with pseudothermal light\cite{RN5,RN6,RN3,RN2} and thereafter with true thermal light\cite{RN7} as well, such as sunlight. Moreover, computational ghost imaging (CGI) was introduced by Shapiro to make this imaging technique full of variety, which keeps the signal beam and exploiting calculated field pattern rather than reference beam\cite{RN8,RN10}. 
Generally, GI is applicable to any wavelength, and has been recently demonstrated with x-rays\cite{RN36,RN43,RN41,RN34}, atoms\cite{RN45}, and even electrons\cite{RN40}. Since GI can have higher resolution beyond the Rayleigh diffraction limit\cite{RN46} and be obtained even in poor illumination\cite{RN42} or turbulent atmosphere\cite{RN39}, it has many potential applications ranging from microscopy\cite{RN47,RN37,RN76} to three-dimensional GI\cite{RN35} to long distance lidar\cite{RN48,RN38} to temporal GI\cite{RN49} and so on. However, no matter what type of ghost imaging method is, the popular way to get the reconstructed image is by a computer imaging algorithm along with a coincidence measurement (photoelectric detection process) between a bucket detection process for signal beam and the known shape of the reference beam. Last year, we proposed a new way to get the reconstructed image with a property of naked-eye imaging\cite{RN18}. 
However, from both the theoretical and experimental results at that time, low contrast image is the main obstacle to push this idea closer to practical applications, since the image is immersed in the reference light beam. Recently, we have solved this problem via all-optical process and the persistence of vision with a special pattern-scanning architecture.

In this letter, we develop a naked-eye ghost imaging method with all-optical computation. In this imaging process, the correlated calculation by the photoelectric coincidence measurement of traditional GI is replaced by all-optical correlation. While the integral imaging process by the computer imaging algorithm of traditional GI  is implemented by the vision persistence effect, where in principle all photosensitive material with the vision persistence effect can be competent for this integral job. 
Meanwhile, to solve the slow imaging speed and low contrast image problems, a low-speed light-modulation disk with a special pattern-scanning architecture is proposed, which also enables high-resolution imaging with lower-order Hadamard vectors and boosts the imaging speed. This light-modulation disk is used to generate a series of light patterns and performs the correlated calculation. At last, the imaging system is tested against moving colored objects, and a high contrast image result is observed directly by eyes. Therefore, our work opens a new way to utilize GI and removes an obstacle to push this idea closer to reality. Besides of optical region, this technique can be applicable for those recently developed GI methods with X-rays, atoms and electrons.

\section{Results}

\begin{figure}[H]
	\centering
	\includegraphics[width=82.5mm]{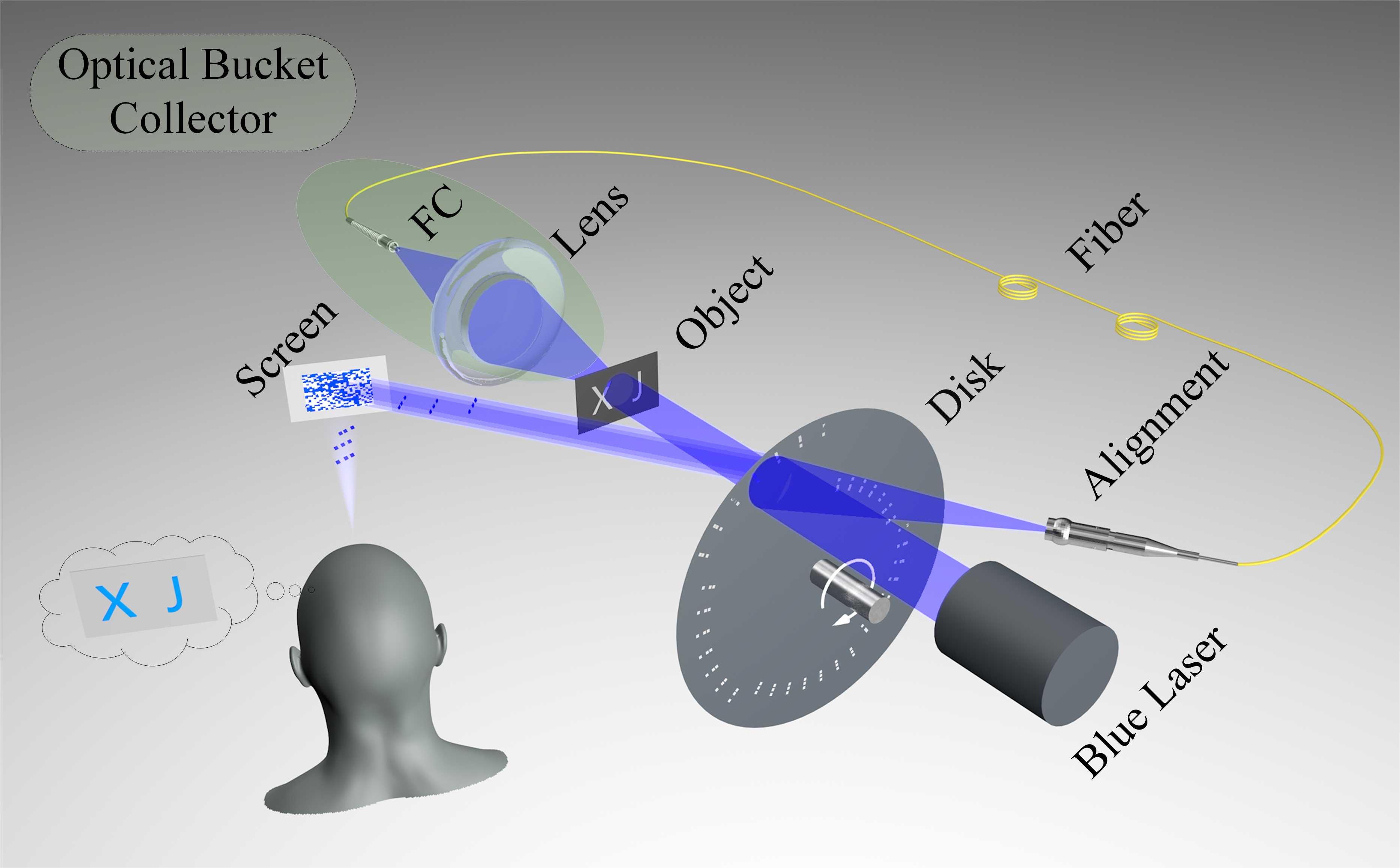}
	\caption{\label{fig01}The original experimental scheme. The high contrast ghost imaging system via the all-optical interaction and the vision persistence effect.}
\end{figure}

The original scheme is shown in Fig {\ref {fig01}}. One blue laser light beam is modulated by a rotating light-modulation disk. Then the modulated light beam illuminates and interacts with moving objects, that are letters ``X'', ``J'', ``T'' and ``U'' with $ 35 \times 35 $ pixels, respectively. The transmitted light after objects is collected by a bucket collector setup with a lens and a FC fiber connector. The collected signal light goes through the fiber and then illuminates the disk overlapping with the blue laser light area. Note that, two light beams propagate in different directions. Here, the correlated calculation is performed by the interaction between the disk and the signal light. Namely, the traditional GI correlated calculation (see Eq. (\ref {GI1})) can be achieved by this setup, 
\begin{equation}\label{GI1}
I(x,y,t) = {{I_1}({x,y},t)} {I_2}(t),
\end{equation}
where ${I_1}({x,y},t)$ stands for the intensity distribution of the instantaneous pattern, ${I_2}(t) $ stands for the corresponding bucket value and $ I(x,y,t) $ is the multiplication of the correlated calculation.  In current setup, the synchronization between ${I_1}({x,y},t)$ and  ${I_2}(t) $ can be realized easily since the light speed is much faster than the rotating speed of disk. Thus, it is also no need to know the shape of the illuminating pattern and the one-to-one match between each pattern and each bucket value. 

The output light ($ I(x,y,t) $) is observed by a photosensitive component such as human eyes, performing integral imaging process, that is
\begin{equation}\label{GI2}
{G^{\left( 2 \right)}}\left( {{x,y}} \right) = \int\limits_t^{t + T} {{I}({x,y},t)} dt,
\end{equation}
where $ T $ stands for the vision persistence time and $ {G^{\left( 2 \right)}}\left( {{x,y}} \right) $ stands for the ghost imaging result.
In this work, we use a CCD camera to mimic the vision persistence effect of human eyes. Since the temporary retention time of human eyes is about 0.02 second in daytime vision, 0.1 second in intermediary vision and 0.2 second in night vision, we choose 0.2 second as the exposure time of CCD. At this point, a high-contrast real-time imaging will be observed by such photosensitive component once the disk rotating at a rate. 

\begin{figure}[H]
	\centering
	\includegraphics[width=62.5mm]{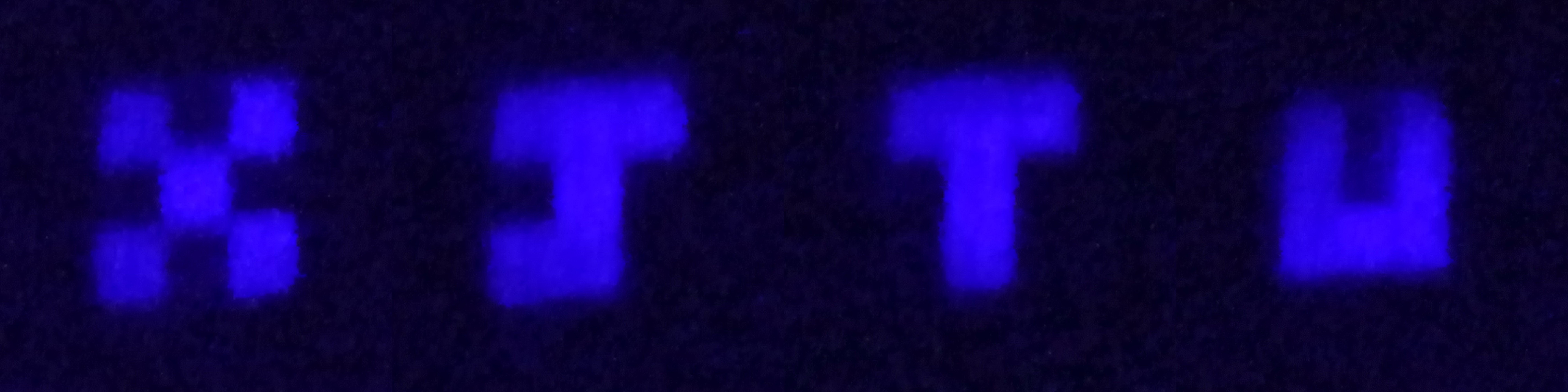}
	\caption{\label{fig02}Ghost imaging results via all-optical process and persistence of vision.}
\end{figure}

Figure {\ref {fig02}} shows the high contrast imaging results (``X, J, T and U'', respectively). Based on our method, the key problem of the image being immersed in the probe light beam is solved. Meanwhile, the bucket photodetector and the computer algorithm for typical GI setup are replaced by the simple all-optical process and the vision persistence effect, which is called all-optical correlation. 

However, this setup has a limitation that the efficiency of bucket collector is very low since the optical fiber coupling is not an easy job. Meanwhile, the lens focal length relies on the light wavelength, which is not suitable for fiber coupling with the multi-color situation. And the optical image filtering effect is also introduced. To realize color ghost imaging, we improve this setup.

\begin{figure}[H]
	\centering
	\includegraphics[width=82.5mm]{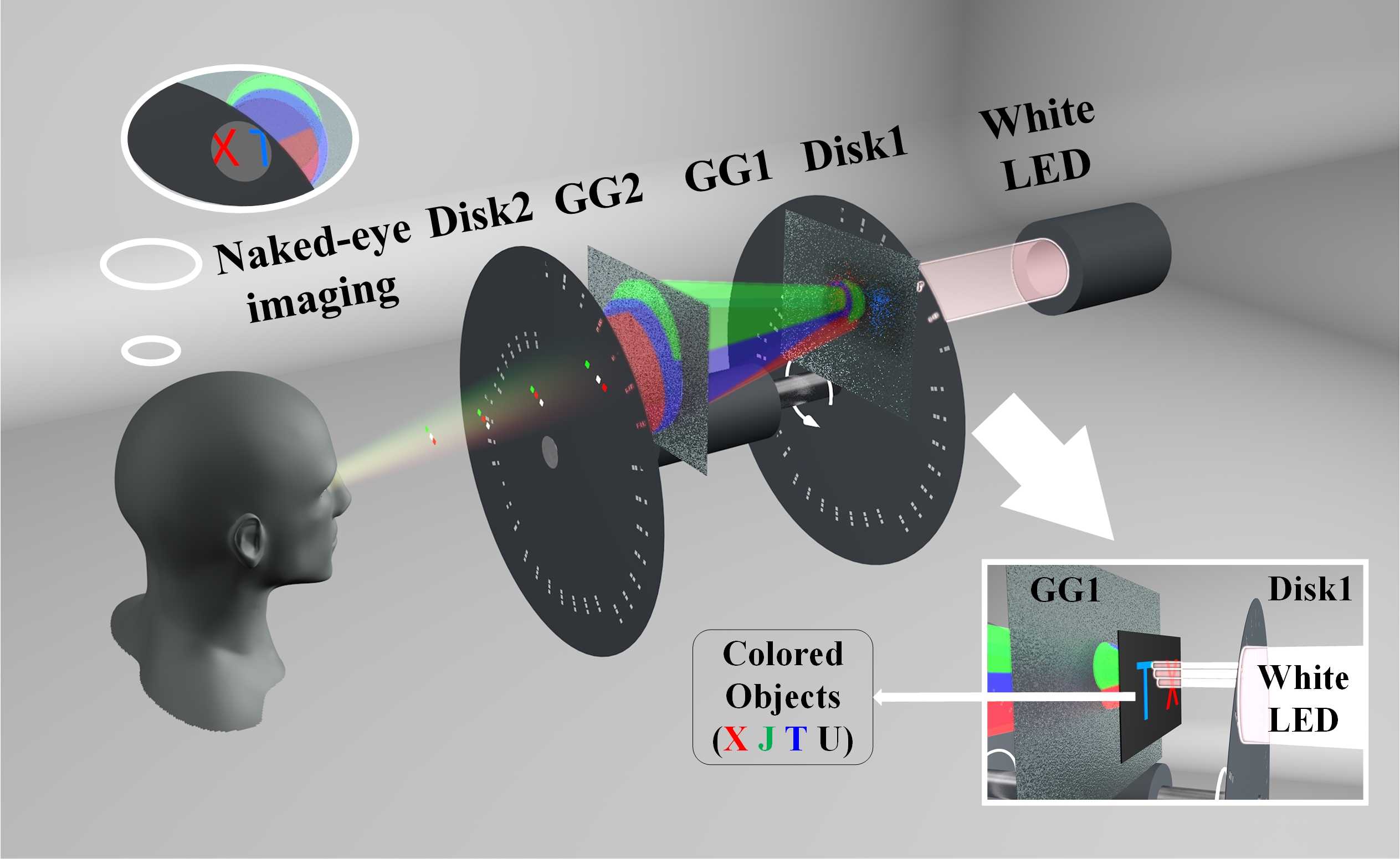}
	\caption{\label{fig03}The scheme of the color ghost imaging system implementation. GG: ground glass.}
\end{figure}
The scheme of the high contrast and color ghost imaging system implementation is shown in Fig. {\ref {fig03}}. One white LED light beam is modulated by a rotating disk (disk 1, same as mentioned in Fig. {\ref {fig01}}). Then it illuminates and interacts with colored objects, which are the red letter ``X'', the green letter ``J'', the blue letter ``T'' and the white letter ``U''  with $ 35 \times 35 $ pixels, respectively. Transmitted light after objects goes through two ground glasses (GG1 and GG2, respectively), which is used to scatter the signal light sufficiently and one can not observe the image directly. Meanwhile, these ground glasses play an optical homogenizer role, which can be understood as a bucket detector in the typical GI setup. Then, this scatted light propagates through the same arranged disk (disk 2) as disk 1, performing the correlated calculation. Finally, the output light is observed by a photosensitive component such as human eyes, performing integral imaging process. Here, two disks are fixed on the same motor, so they are rotating at the same rate. We also use the CCD camera to mimic the vision persistence effect of human eyes. By this setup, a high contrast and color image will be observed once two disks rotating at certain rate. 
\begin{figure}[H]
	\centering
	\includegraphics[width=62.5mm]{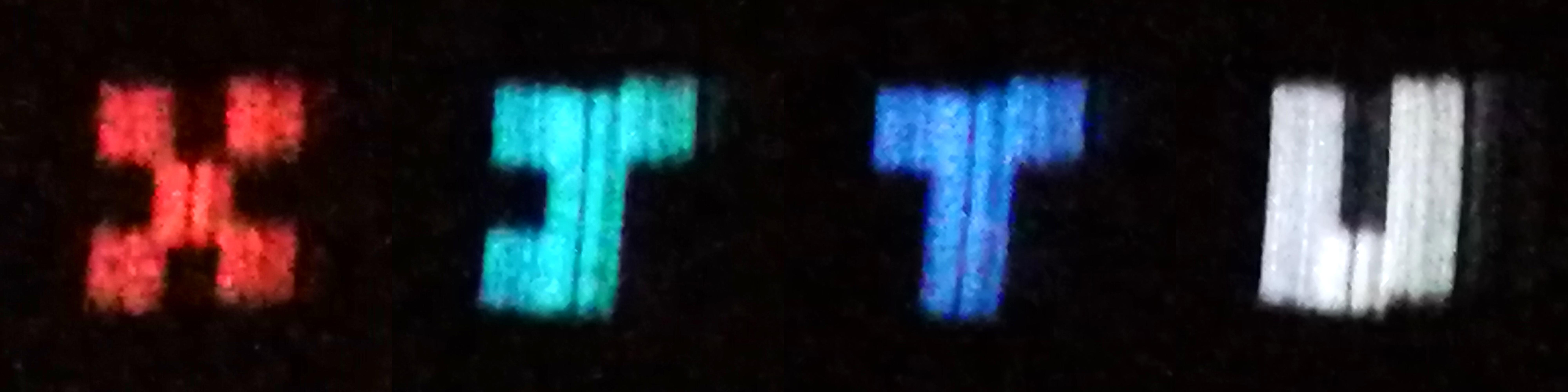}
	\caption{\label{fig04}Color ghost imaging results via all-optical process and persistence of vision. }
\end{figure}
Figure {\ref {fig04}} shows the high contrast and color imaging results (``X, J, T and U'', respectively). Especially, high-contrast real-time imaging video for moving colored objects is shown in the supplement(see Visualization 1). Based on the improved method, the limitation mentioned above  is solved.  

In the whole idea, the key component is the specially designed light-modulation disk. It is not only used to generate a series of modulated light patterns and does the correlated calculation, but also provides better quality image results with high contrast and high resolution.
In the method section, we will show how to realize this high contrast and color ghost imaging process.

In summary, high-contrast real-time imaging for moving colored objects is realized by all-optical computation. The bucket detecting process with the photoelectric way and correlated calculation, as well as integral imaging for traditional GI, are replaced by our new way. The obstacle to realizing high-contrast real-time imaging for moving colored objects is removed by a special pattern-scanning architecture. Meanwhile, high resolution and the boosted imaging speed can be obtained with low pixel illumination from a low-speed rotating light-modulation disk. This work opens a new way to utilize GI, which can be used to 3D GI visualization, GI virtual reality and so on. 

\section{Materials and Methods}
This special pattern-scanning architecture on a low-speed light-modulation disk realizes two functions. Firstly, it is used to generate a series of light pattern and does the correlated calculation. Secondly, its special design ensures the high contrast imaging results. Now we show this idea step by step. 

\subsection{Imaging analysis.} We define that $N$ is the pattern pixel number. Ghost imaging can be expressed as follow
\begin{equation}
G^{(2)}=\frac{1}{m}AY\sim AY=AA^{T}X \sim Cov(A)X,
\label{F1}
\end{equation}
where \textit{m} is the  numbers of sampling, \textit{A} is the measurement matrix, \textit{Y} is the output value of the bucket detector, $X$ is the object.  For convenience, we denote $Cov(A)$ as $C$, 
\begin{equation}
C_{ij}=
\begin{cases}
C_{min}& i\neq j\\
C_{max}& i=j
\end{cases}.
\label{F2}
\end{equation}

So, one can get the imaging contrast from the definition of Eq. (\ref {Vis})
\begin{equation}
\begin{split}
Contrast&=\frac {Max-Min}{Max+Min},\\
Max&=C_{max}+(N_{obj}-1)C_{min},\\
Min&=N_{obj}C_{min},
\end{split},
\label{Vis}
\end{equation}
where we set object matrix element as $N_{obj}$ pixels are 1 and the others is 0.

\subsection{Speckle structure design and analysis.} 
For a class of objects to be imaged, getting a suitable speckle pattern for higher contrast and fewer speckles in this system remains a challenging task. 
Usually, the traditional artificially designed speckle pattern (Hadamard) is better than random speckle pattern, since fewer speckle's number is needed with the same contrast obtained. Hadamard pattern's number depends on the total pixel's number of object we imaging. Therefore, to get a higher contrast imaging result, the way to redesign the structure of Hadamard is a feasible plan.  

\emph{\textbf {Step 1: Reduction of the matrix.}} The Hadamard measurement matrix ($N_{H}\times N_{H}$) is a  special kind of matrix, in which $N_{H}$-by-$N_{H}$ Hadamard matrix with $N_{H} > 2$ exists only if the remainder of $N_{H}$ after division 4 is equal to 0, where $N_{H}$, $N_{H}/12$, or $N_{H}/20$ is a power of 2 \cite{RN14,RN11}.
When one takes the negative to zeros for the projection, one can figure out that the first pattern and every first pixel of each pattern are 1. It is little significance for projecting the invariant pixels and projecting the speckle with a total of 1 to measure the detail of the object. In addition, it will increase the imaging background, leading to a decrease of the imaging contrast. So, it makes sense to abandon these pixels and speckle. Therefore, by abandoning the first row and the first column, the Hadamard measurement matrix will become $(N_{H}-1)\times (N_{H}-1)$.
For example, the original Hadamard matrix of order 8 is shown as
\begin{equation}
H=\left[                 
\begin{array}{rrrrrrrr}
1  &  1  &   1  &   1  &  1  &   1  &  1 &  1 \\
1  & -1  &   1  &  -1 &  1  &  -1  &  1 & -1\\
1  &  1  &  -1  &  -1 &  1  &   1  & -1 & -1\\
1  & -1  &  -1  &   1 &  1  &  -1  & -1 &  1\\
1  &  1  &   1  &   1  &-1  &  -1  & -1 & -1\\
1  & -1  &   1  &  -1 & -1  &  1   & -1 &  1\\
1  &  1  &  -1  &  -1 & -1 &  -1   &  1 &  1\\
1  & -1  & -1   &  1  & -1  &  1   &  1 & -1\\ 
\end{array}\right ] _{8\times8}.
\label{F10}
\end{equation}
Then the reduced Hadamard matrix of order 8 is become
\begin{equation}
\widehat {H}=\left[                 
\begin{array}{rrrrrrr}
0  &  1  &   0  &   1  &  0  &   1  &  0  \\
1  & 0  &   0  &  1 &  1  &  0 &  0 \\
0  &  0  &  1  &  1 &  0  &   0 & 1 \\
1  & 1  &  1  &   0 & 0  &  0  & 0 \\
0  &  1  &   0  &  0  &1  &  0  & 1\\
1  & 0  &  0  &  0 & 0  &  1   & 1 \\
0 &  0 &  1  &  0 & 1 &  1   & 0 \\ 
\end{array}\right ] _{7\times7}.
\label{F11}
\end{equation}
One can use 7 patterns reshaped from a 7-row vector of the matrix to get image. 
As a result, the operation will contribute to the contrast $C_{ij}$ as
\begin{equation}
C_{ij}=
\begin{cases}
1& i\neq j\\
3& i=j
\end{cases}.
\label{F12}
\end{equation}
And the background will be half of the peak when using the small hole as an object. The contrast is up to 1/2. 
In this way, the relationship between contrast and resolution is obtained. One can get the expression of $C_{ij}$
\begin{equation}
C_{ij}=
\begin{cases}
\frac{N+1}{4}-1& i\neq j\\
\frac{N+1}{2}-1& i=j
\end{cases},
\label{F13}
\end{equation}
where $N+1$ is the order ($N_{H}$) of the Hadamard matrix, and the pattern pixel's number is $N$ with $N\geq3$. So, the object with which $N_{obj}$  pixels are 1 and the other are 0 is used for imaging, one can get the equation of contrast via Eq. (\ref {F13})
\begin{equation}
contrast_{\widehat {H}}=\frac{1+N}{1+N+2N_{obj}(N-3)}.
\label{F14}
\end{equation}
From the Eq. (\ref {F14}), one can get that $contrast_{\widehat {H}}$ increases with decreasing $N_{obj}$. So, this method can improve the contrast ratio of small duty ratio area imaging. Moreover, the pattern reshaped from the reduction Hadamard matrix will remain only one lighted pixel when $N=3$, which is similar to the laser point scanning technique.

\begin{figure}[H]
	\centering
	\includegraphics[width=62.5mm]{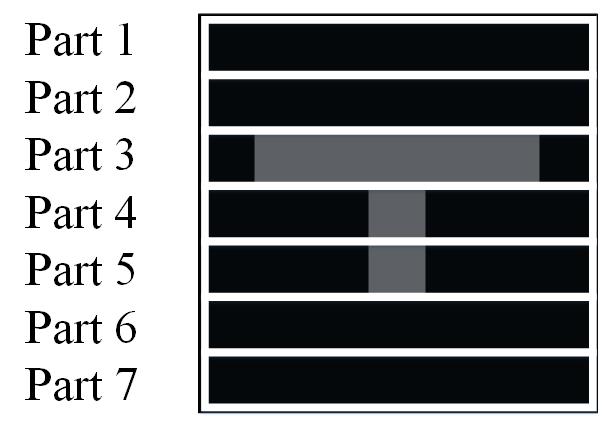}
	\caption{\label{fig05}Imaging partition in the first dimension. }
\end{figure}

\emph{\textbf {Step 2:  Imaging partition in the first dimension.}} As shown in Fig. {\ref {fig05}}, there is a way that we divide the object ($n\times n$) into $n$ row part with each part is $1\times n$. It can increase local contrast in imaging, and the effect of detail enhancement is achieved. 
Here, the part contrast is
\begin{equation}
\begin{split}
contrast_{\widehat {H}\_Part}&=\frac{1+N_{Part}}{1+N_{Part}+2N_{obj}(N_{Part}-3)},\\
N_{Part}&=\sqrt {N}=n.
\end{split}
\label{F16}
\end{equation}
Generally, digital micromirror device (DMD) can realize this work. One can project the $1\times n$ Hadamard pattern $n$ times to the row part of the object for this part imaging, step by step. When the last row part of the object is finished, the imaging result can get. 

\begin{figure}[H]
	\centering
	\includegraphics[width=62.5mm]{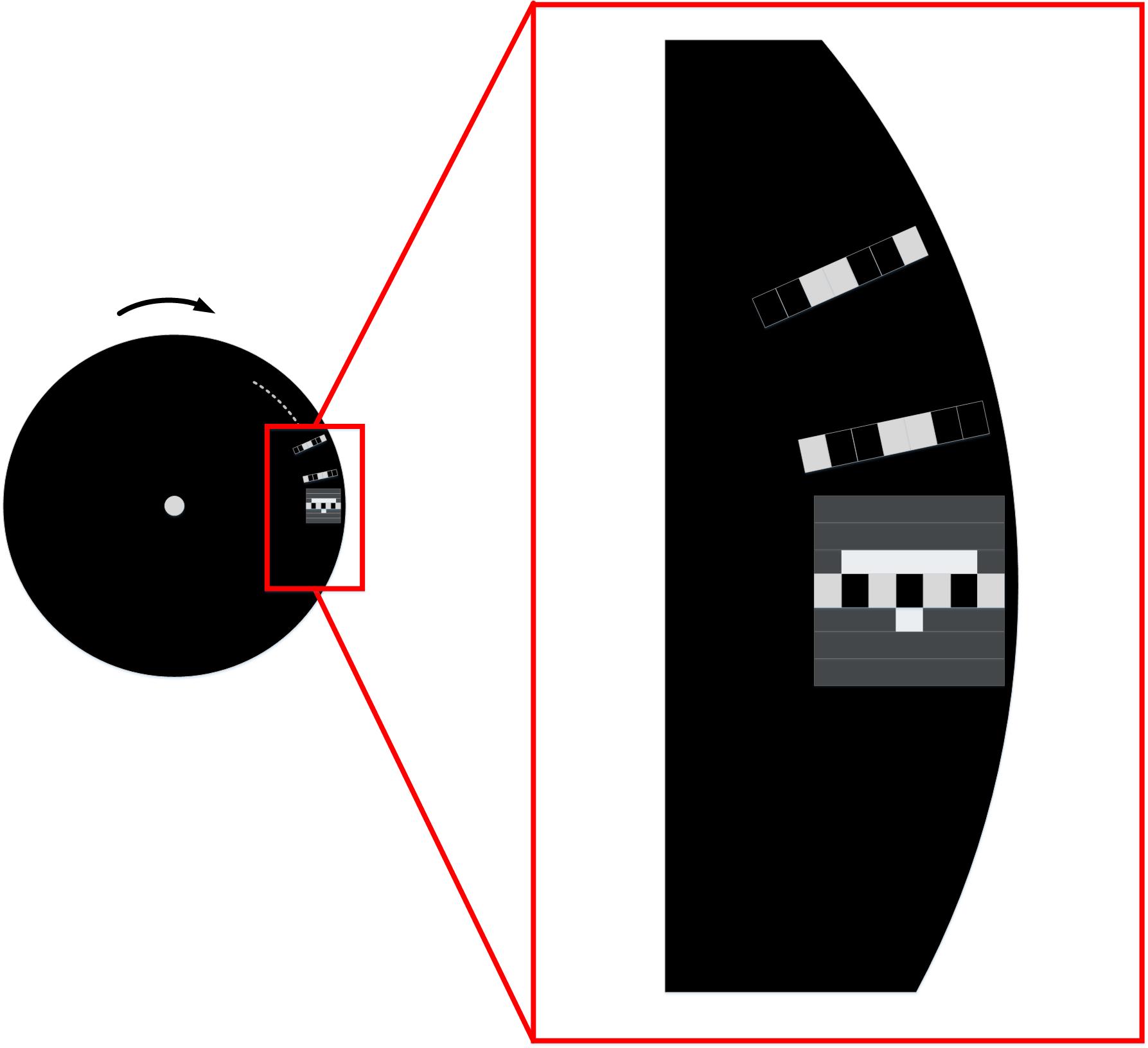}
	\caption{\label{fig06}A rotating mask with a pattern hole. }
\end{figure}

\emph{\textbf {Step 3:  New projection method.}} The projection method is a simple to understand. However, we can change the projection order that one can use the first $1\times n$ Hadamard to go through each row part of object top to bottom, then use the remaining $1\times n$ Hadamard to do the same thing. We call it as pattern moving. So, one can improve resolution by reducing the step length. 
When it comes to the pattern moving, this is a new projecting method, which is better than the DMD method we mentioned above. A new projecting method is a rotating mask, which is a disk with a pattern hole on the edge as shown in Fig. {\ref {fig06}}. It is a low-cost device that can be forming very quickly. In addition, there are $n$ mask to imaging $n\times n$ object. This method can greatly reduce the number of the mask.

\begin{figure}[H]
	\centering
	\includegraphics[width=82.5mm]{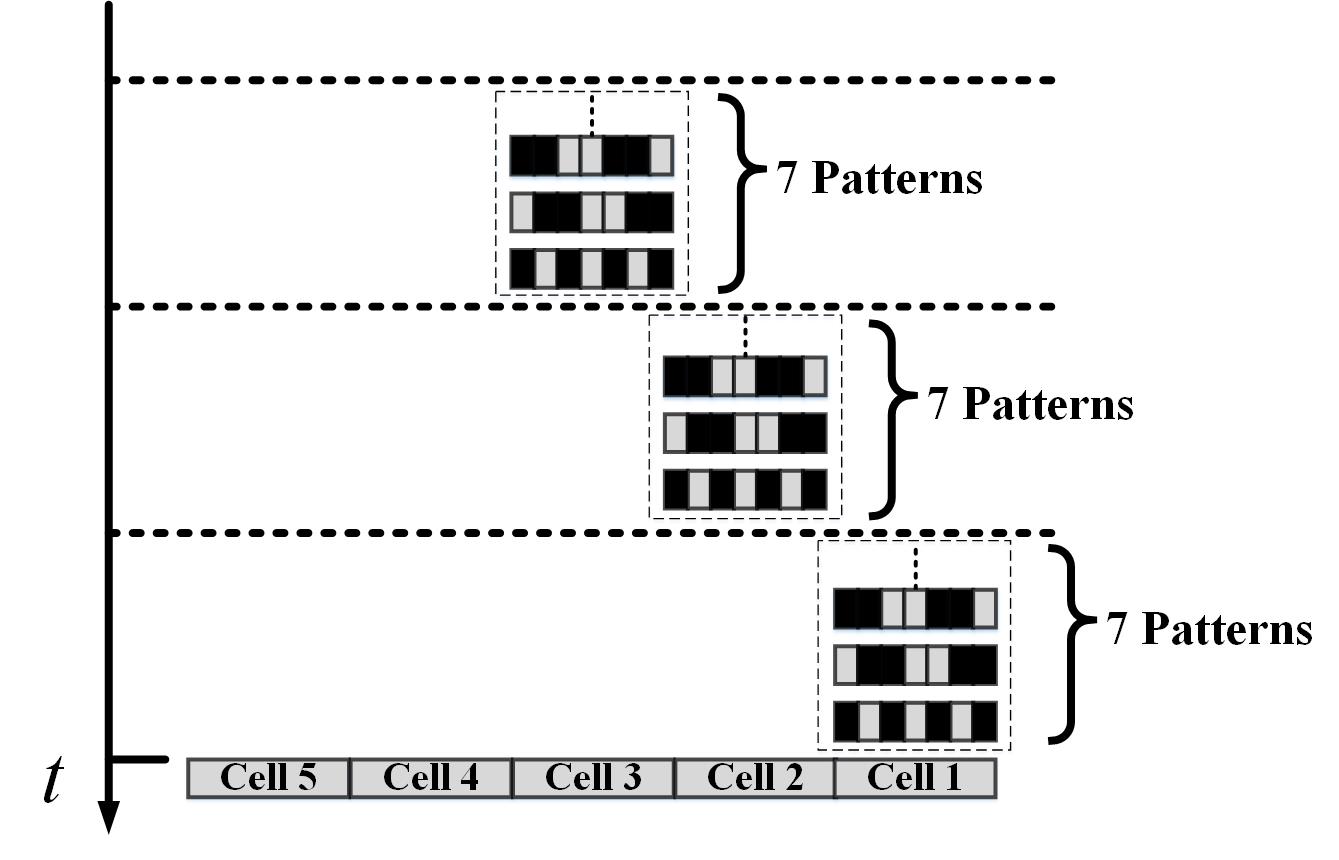}
	\caption{\label{fig07}Cell structure. }
\end{figure}

\emph{\textbf {Step 4: Imaging partition in the second dimension.}} 
In order to get a higher contrast, one can realize the imaging partition in the second dimension. One can divide the row part object ($1\times n$) into $k$ row cell with each small cell be $1\times (n/k)$ as shown in Fig. \ref {fig07}. Instead of using the $1\times n$ Hadamard pattern moving, we use the Hadamard cell moving, which consists of a complete set of Hadamard pattern. Its contrast can be expressed as

\begin{equation}
\begin{split}
contrast_{\widehat {H}\_Cell}&=\frac{1+N_{Cell}}{1+N_{Cell}(2N_{Cell}-5)},\\
N_{Cell}&=\frac{\sqrt {N}}{k}.
\end{split}
\label{F17}
\end{equation}

In order to get high contrast via Hadamard pattern, apart from the sample scanning, it is a suitable choice that one can take $N_{Cell}=7$. So, we adopt $ n=35 $, $ k=5 $, and $N_{Cell}=7$ to realize imaging.

\begin{acknowledgments}
We wish to acknowledge the support of National Basic Research Program of China (973 Program) (Grant No. 2015CB654602); Key Scientific and Technological Innovation Team of Shaanxi Province (Grant No. 2018TD-024); 111 Project of China (Grant No. B14040).

\end{acknowledgments}

\section*{REFERENCES}
\bibliography{VGI}

\end{document}